# Mobile Internet from the Heavens

Farooq Khan
Samsung Electronics
Richardson, Texas, USA

*Abstract* — Almost two-thirds of the humankind currently does not have access to the Internet, wired or wireless. We present a Space Internet proposal capable of providing Zetabyte/ month capacity which is equivalent to 200GB/month for 5 Billion users Worldwide. Our proposal is based on deploying thousands of low-cost micro-satellites in Low-Earth Orbit (LEO), each capable of providing Terabit/s data rates with signal latencies better than or equal to ground based systems.

## I. INTRODUCTION

Mobile connectivity has transformed daily life across the globe becoming one of the most dramatic game-changing technologies the world has ever seen. As more people connect to the Internet, increasingly chat to friends and family, watch videos on the move, and listen to streamed music on their mobile devices, mobile data traffic continues to grow at unprecedented rates. Actually, this surge in demand is following what we can informally call an omnify principle. Here, *omnify* stands for *Order of Magnitude Increase every Five Years*. Which means, demand for data increases 10 times every 5 years and will continue to increase at this pace with expected 1,000 times increase in the next 15 years. This increase in demand is similar to the memory and computing power growth following Moore's law which offered a million fold more memory capacity and the processing power in the last 30 years. For wireless communications, it is more appropriate to measure advances in 5 years and 10 years timeframe as a new generation wireless technology is developed every 10 year and a major upgrade on each generation follows 5 years afterwards as shown in Figure 1. Global mobile traffic already surpassed 1 Exabyte/ month mark in 2013 and is projected to grow 10 fold exceeding 10 Exabytes/ month within 5 years in 2018 [1]. With this trend, in 2028 global mobile traffic will cross the 1 Zetabyte/ month which is equivalent to 200 Gigabytes/month for 5 Billion users. Wi-Fi offload accounts for almost as much traffic as on mobile networks and also follows a similar growth trend with another Zetabyte/ month mobile traffic flowing over Wi-Fi networks by 2028. We also expect peak wireless data rates to follow a similar trend, increasing from 1 Mb/s in year 2000 with 3G to around 10 Gb/s with 5G in 2020 and finally 1 Tb/s with 6G in 2030 offering million times increase in 30 years as depicted in Figure 1. The peak data rates for Wi-Fi follow a similar trend with a few years lead. This means wireless data is catching up with the memory, storage and computing capabilities that are already available to deal with these massive amounts of data.

Traditionally, all wireless communications, with the exception of point-to-point microwave backhaul links, used a relatively narrow band of the spectrum below 3GHz. This sub-3GHz spectrum has been attractive for Non-Line-of-Sight (NLOS) point-to-multipoint wireless communications due to its favorable propagation characteristics. The large antenna aperture size at these frequencies enables broadcasting large amount of power which allows signals to travel longer distances as well as bend around and penetrate through obstacles more easily. This way, the sub-3GHz spectrum allowed providing wide area coverage with a small number of base stations or access points (APs). However, with mobile data traffic explosion, modern wireless systems face capacity challenges requiring deployment of more and more base stations with smaller coverage area. However, the number of small cells that can be deployed in a geographic area is limited due to costs involved for acquiring the new site, installing the equipment and provisioning backhaul. In theory, to achieve 1,000-fold increase in capacity, the number of cells also needs to be increased by the same factor. Therefore, small cells alone are not expected to meet the capacity required to accommodate orders of magnitude increase in mobile data traffic demand in a cost effective manner.

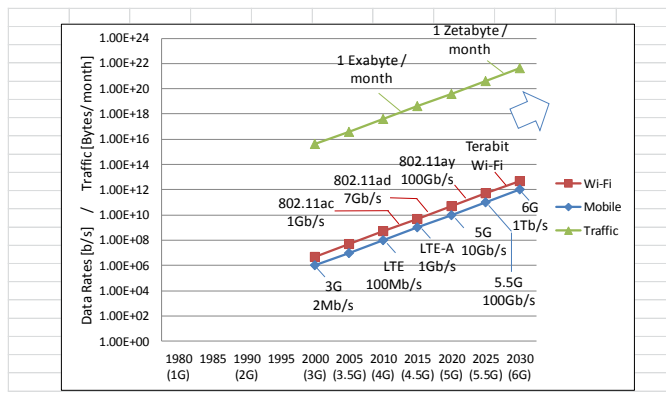

**Figure 1 Wireless data follows an omnify principle**

In order to address the continuously growing wireless capacity challenge, the author and his colleagues pioneered use of higher frequencies referred to as millimeter waves with a potential availability of over 100GHz spectrum for 5G mobile communications [2]-[5]. At millimeter wave frequencies, radio spectrum use is lighter and very wide bandwidths along with a large number of smaller size antennas can be used to provide orders of magnitude increase in capacity needed in the next 15 to 20 years. Moreover, in order to connect the remaining two-thirds of the humankind that currently does not have access to the Internet, we will need to complement cellular and Wi-Fi networks with satellites and other aerial systems such as those using unmanned aerial vehicles (UAVs) [6].

Traditionally, geostationary satellites have been preferred for communication due to several advantages they present. Firstly, the entire planet, with the exception of small circular regions centered at the north and south geographic poles, can be covered with as little as three such satellites, each separated by 120 degrees of longitude. This is because a single geostationary satellite is on a line of sight with about 40 percent of the earth's surface. Secondly, a geostationary satellite can be accessed using a directional antenna, usually a small dish, aimed at the spot in the sky where the satellite appears to hover and then left in position without further adjustment.

A key problem with geostationary satellites at altitude of 35,786Km, however, is that radio signals take approximately 0.25 of a second to reach and return from the satellite. This signal delay can undermine performance of interactive services such as voice / video conversations, gaming and other applications running on common network protocols such as TCP/IP. Other disadvantages of geostationary satellites are lower overall capacity due to larger coverage footprint (reduced frequency reuse) and higher propagation loss.

In this paper, we discuss a Low Earth Orbit (LEO) satellite system capable of carrying a total at least one Zetabyte/ month data traffic by employing thousands of high capacity micro-satellites, each operating at Tb/s or higher data rates. The so called micro-satellites in LEO orbit use less power due to proximity to Earth, are smaller in size, lower in weight, and therefore are easier to launch.

I. SPACE INTERNET

A major barrier to expanding connectivity to the communities that currently does not have access to the Internet is the cost of providing the service. With the goal to reduce cost, the author proposed a new wireless architecture referred to as multi-comm-core (MCC) that can scale to Terabit/s data rates for ground based local area and wide area wireless access, for wireless backhaul as well as access via unmanned aerial vehicles (UAVs) and satellites [6].

With our goal of Zetabyte/ month data capacity, each satellite in the system would need to provide Terabit/s or higher data arte. An example of such a system that we refer to as the *Space Internet* is depicted in Figure 2. Here, all three types of links in the system namely uplink from ground station to satellite, inter-satellite link and downlink from satellite to ground or air operate at data speeds in excess of Tb/s. Note that uplink and downlink in this context refers to use of different frequency spectrum for these links. From data traffic perspective, these links are bi-directional carrying traffic in both directions. We remark that in many cases, satellite ground stations can be co-located with large data centers reducing overall latency and cost of the system. We will later see in the link budget calculations that antenna array size and power consumption may be too large for handheld mobile devices to directly communicate with the satellites. In this case, satellites provide the backhaul to the cellular base station or Wi-Fi access point which in turn provide connectivity to the handheld mobile devices. However, for communication between satellites and airplanes / vehicles, it may be possible to mount large antenna arrays with enough signal gain enabling direct communication with the satellites.

We assume that the satellite itself and the systems on the ground and the air directly communicating with the satellite such as ground stations (data centers), cellular and Wi-Fi access points, airplanes and vehicles will use phased array antennas. A key benefit of phased arrays is that they allow steering the beam with the flexibility and speed of electronics rather than with much slower and less flexible mechanical steering used for traditional parabolic dish antennas. Moreover, with phased arrays, a satellite can form multiple beams to spatially separated systems on the ground and in the air. Similarly, a ground station can communicate with several satellites simultaneously by generating multiple independent beams and scanning them in the two orthogonal planes.

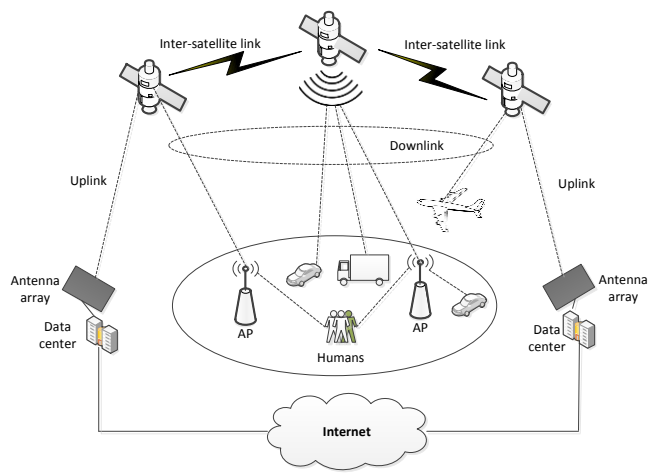

**Figure 2 Space Internet system**

*A. Spectrum*

Traditional satellite communications and wireless backhaul markets have been highly fragmented due to lack of standards and use of proprietary technologies. One reason for this fragmentation is use of lower frequencies (below 6GHz) for local area and wide area wireless access while wireless backhaul and satellite communications using higher millimeter wave spectrum. However, with the 5G vision of providing wireless access in the millimeter wave spectrum [2]-[4], a single standard-based wireless technology can be developed for access, backhaul and satellite communications eliminating fragmentation and thereby reducing costs of providing wireless services. With the use of higher millimeter wave frequencies, back-haul links and access can share the same spectrum due to highly directional nature of beamformed millimeter wave transmissions. However, for satellite and /or UAV communications, methods to avoid or cancel interference needs to be considered when ground based and aerial (satellites, UAVs) systems share the same spectrum. In most cases, ground based and aerial systems should be able to use non-overlapping frequencies due to vast spectrum available at the millimeter waves.

Space Internet satellites providing Tb/s data rates would require tens of GHz of spectrum which would require use of spectrum across wide frequency range. In Table 1, we list spectrum bands between 10 and 275 GHz that are currently designated for satellite communications in the FCC frequency allocation chart [7]. We note total spectrum availability of over 100GHz which includes 57.75GHz for uplink, 56.2GHz for downlink and 38.75GHz for inter-satellite communication. In some cases, the same spectrum band such as 66-71GHz band is shared between downlink and inter-satellite links. Moreover, 28GHz band is allocated for LMDS as primary and the satellite uplink on a secondary basis. Also, 59-66GHz oxygen-absorption band designated for inter-satellite link is also an unlicensed band used for short-range connectivity and point-to-point broadband wireless. In this case, inter-satellite links are not expected to interfere with ground based systems due to high absorption in the atmosphere.

**Table 1 Spectrum for satellite communications**

| Uplink | | Downlink | | Inter-Satellite | |
|---|---|---|---|---|---|
| Frequency [GHz] | BW [GHz] | Frequency [GHz] | BW [GHz] | Frequency [GHz] | BW [GHz] |
| 12.5-13.25 | 0.75 | 10.7-11.7 | 1.0 | 22.55-23.55 | 1.0 |
| 13.75-14.8 | 1.0 | 17.7-21.2 | 3.5 | 25.25-27.5 | 2.25 |
| 27.5-31.0 | 3.5 | 37.0-42.5 | 5.5 | 59.0-66.0 | 7.0 |
| 42.5-47.0 | 4.5 | 66.0-76.0 | 10.0 | 66.0-71.0 | 5.0 |
| 48.2-50.2 | 2.0 | 123.0-130.0 | 7.0 | 116.0-123.0 | 7.0 |
| 50.4-51.4 | 1.0 | 158.5-164.0 | 5.5 | 130.0-134.0 | 4.0 |
| 81.0-86.0 | 5.0 | 167.0-174.5 | 7.5 | 174.5-182.0 | 7.5 |
| 209.0-226.0 | 17.0 | 191.8-200.0 | 8.2 | 185.0-190.0 | 5.0 |
| 252.0-275.0 | 23.0 | 232.0-240.0 | 8.0 | | |
| Total | **57.75** | Total | **56.2** | Total | **38.75** |

*B. Latency*

A common misconception about satellite communications has been that signal delays are always higher compared to ground based systems. As discussed earlier, this is only true for satellites at very high altitudes such as satellites in geosynchronous orbit. It is also generally true that for communication between two points on earth, a signal going through space has to travel longer distance. However, the speed of signal propagation is generally about 1.4 times faster in space or air compared to propagation in a fiber optic cable with refractive index of $n = 1.4$ as depicted. Let us calculate the satellite altitude $r_a$ at which the signal delay $t_p$ is the same for space and ground based systems as below:

$$t_p = \frac{2r_a + 2\pi q(r + r_a)}{C} = \frac{2\pi q r}{C/n}$$

Where $q$ is the fraction of the earth circumference (40,075 Kms) that the signal has to travel, $r$ is the radius of earth and $C$ is the speed of light in free space. By rearranging the terms in the above equation we can write the satellite altitude $r_a$ as below.

$$r_a = \frac{(n-1)r}{\left(1 + \frac{1}{\pi q}\right)}$$

We plot satellite altitude $r_a$ as function of distance measured in terms of fraction of Earth's circumference in Figure 3. For example to go half-way around earth ($q = 0.5$), satellites deployed at altitude of 1,557 Kilometers would provide the same one-way signal delay of around 93.5ms as a fiber optic cable. Here the signal in space would need to travel approximately 40% farther (28,021 Kilometers) compared to a fiber optic cable (20,037 Kilometers). However, the signal delay for the two systems is the same as signal propagates 40% faster in free space than in an optical fiber. For the same case, with satellite altitude less than 1,557 Kilometers, satellite based system would offer lower signal delay compared to transmission through a fiber optic cable. This observation is true for long-distance communication. For short-distance communication, ground based systems may provide lower latency because in this case the signal delay in reaching and returning from the satellite dominate the overall delay. However, for short-range communication, the overall delays may, anyway, be acceptable with signal round trip time to a LEO satellite of just around 20ms. In practice, the overall latency can be higher due to signal processing and signal amplification delays both in space based and optical fiber based systems. We also note that in general, to keep signal delays the same as in ground based systems, satellites would need to be deployed in LEO orbit below 2,000 Kilometers altitude. Another advantage of satellites over ground based communication is that traffic can be routed dynamically, from one satellite to the other finding the shortest path to the destination and hence reducing signal latency.

*C. Constellation*

As discussed, by year 2028, both cellular and Wi-Fi will be carrying data traffic in excess of one Zetabyte/ month. Our goal here is to design a Space Internet with similar capacity. This space Internet can then provide back-haul for cellular and Wi-Fi as well as direct communication with the satellite connecting the world's population currently without Internet access. With the satellite-based backhaul, cellular and Wi-Fi deployments become practical in remote regions of the earth where there is no wired Internet infrastructure. We like to keep the number of satellites as small as possible with each satellite providing ultra-high capacity in excess of Tb/s. We note that Zetabyte/ month data capacity is equivalent to 3,200 links at 1Tb/s. With our Tb/s satellites and assuming 2/3 utilization of satellite capacity, we will require about 4,600 satellites to carry Zetabyte/ month data traffic. With this large number of satellites needed to meet the goal, our motivation for low-cost micro-satellites with lower development and launch costs become even more apparent.

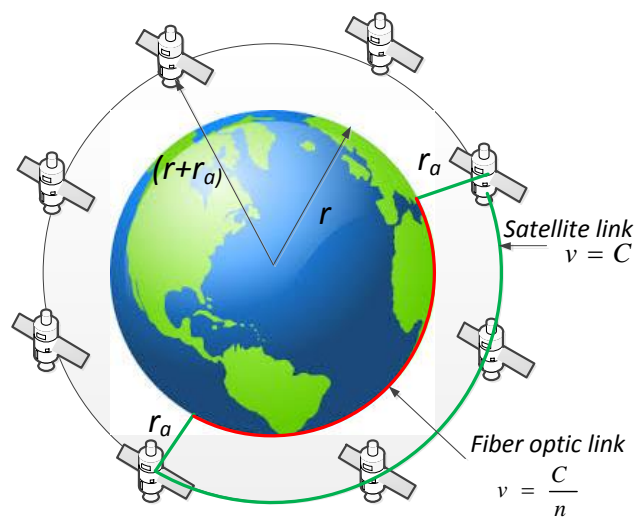

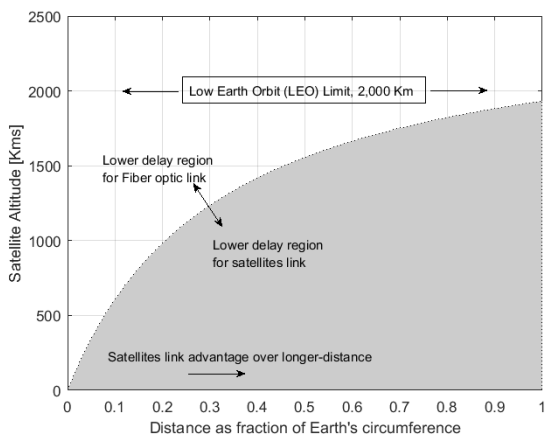

**Figure 3 Satellite altitude for equal delay between optical fiber and space based communication systems**

Figure 4 which allows natural scaling of the bandwidth and the hardware for satellites and their ground/air counterparts. The MCC architecture is attractive from power consumption perspective when ultra-fast data rates in the tens or hundreds of Gigabit/s to Terabit/s range require many GHz of bandwidth. This architecture is equally applicable to other wireless systems evolving to these data rates such as ground based local area (Wi-Fi) and wide area wireless (mobile) access, for wireless backhaul as well as access via unmanned aerial vehicles (UAVs) [6]. The motivation for a transition to MCC architecture is the same as transition to multi-core architecture for computing in the past 10 years which was mainly driven by unsustainable level of power consumption implied by clock rates in excess of GHz. We expect a similar transition to happen in wireless communications to achieve the targeted ultra-fast data rates with reasonable complexity and power consumption.

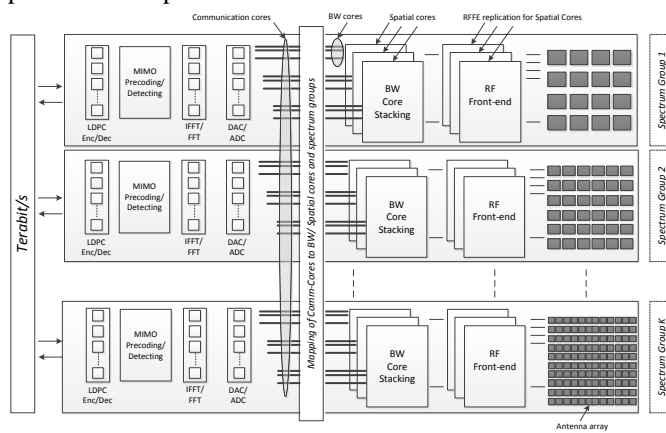

**Figure 4 Multi-Comm-Core (MCC) radio architecture**

Unlike geostationary satellites, LEO satellites do not stay at a fixed location in the sky making it necessary for the ground stations to continuously track them when they move in the sky. The orbital period for LEO satellites range from about 88 minutes at 160 Kilometers altitude to about 127 minutes at 2,000 Kilometers altitude. This means a given satellite is going to be visible to the ground systems for a relatively short period. Therefore, an entire constellation of satellites would be necessary to maintain constant coverage where a new satellite shows up at the location of a previous satellite which has moved out of sight. In addition to higher overall capacity, another advantage of a larger constellation size is that it allows us to further optimize satellite positions in the constellation to provide relatively higher capacity over certain high density geographic areas.

*D. Radio Architecture*

As discussed earlier, a viable radio architecture for the Space Internet is the multi-comm-core (MCC) approach shown in

We will see in the next section that to achieve ultra-fast Tb/s data rates, our system would need to operate at bandwidths of 32 GHz or higher. In practice, these high bandwidths may not be available in a contiguous manner (see Table 1) even in the higher millimeter waves or the RF front-end and the antennas may not support such high bandwidths with required efficiency. Therefore, we expect the system to use a set of RF front-end and antenna arrays covering each spectrum group of frequencies as shown in Figure 4. Within each spectrum group, multiple bandwidth (BW) cores each supporting 1-2GHz bandwidth and are stacked. The RF front-end and BW core stacking blocks within each spectrum group are replicated to provide spatial cores (to support multi-beam and/or MIMO capability). Each BW core has its own set of data converters (ADC/ DACs), FFT/IFFT, channel coding and other baseband functions and these blocks/functions are replicated across spatial cores. The MIMO/beam processing is done jointly across the spatial cores within the same BW core.

The proposed architecture allows us operating in more energy efficient way using smaller blocks operating at lower clock frequencies and using multiple of them together to achieve the high data rate. It also provides high degree of flexibility to achieve a given data rate through use of various combinations of BW cores, spatial cores and spectrum groups based on HW and spectrum availability. Moreover, it allows scaling down the HW to meet given cost, form factor, power consumption or complexity requirements.

## II. LINK BUDGET ANALYSIS

We carry out the link budget analysis for a Tb/s satellite link in Table 2. We first calculate the data rate per comm-core for a reference system with 1GHz core bandwidth at 100 GHz frequency and a range of 1,500 Kilometers. We note that with a total of 256 comm-cores, we can reach 1.21 Terabit/s data rate. This can, for example, be achieved with 32 BW cores with total 32 GHz bandwidth and 8 MIMO cores or beams or other combinations of BW cores and MIMO cores. This means that very large bandwidths and large number of MIMO beams will be required to achieve Terabit/s data rates. With the multi-comm-cores architecture, the total bandwidth (32GHz in this example) can be aggregated across different spectrum bands given in Table 1. Another observation is large amount of total transmit power which will be 512 Watts for 256 cores just for the power amplifiers only with the assumption of 2W (33dBm) of power per core. Accounting for power amplifier efficiency and consumption in other RF components and the baseband, we expect the total power consumption in the few KW range or higher. This type of power is feasible for the ground stations communicating with the satellite(s) and for transmissions from satellites to cellular and Wi-Fi access points, airplanes and vehicles. However, transmissions from cellular and Wi-Fi access points, airplanes and vehicles to the satellites may not need to use all the 256 cores thereby reducing the total power consumption. This means the aggregate data rate of Tb/s from/to the satellite in this case will be shared among multiple access points on the ground and the air in the coverage area of the satellite.

The trasmit antenna total gain of 53dBi can be achieved with various combinations of antenna element gain and antenna array size based on the requirements on the desired beem steering range and capabilities. An example configuration can be an antenna array size of 1024 elements (30 dB) with antenna elements gain of 23dBi. In general, we can assume larger element gain for satellite communication comapred to ground based systems because the possible range of signal directions is limited in satellite systems. With antenna arrays used both at the transmitter and the receiver, higher frequencies will benefit from higher array gain for the same total antenna area due to smaller wavelengths [1][3]. This will help compensate for other losses that tend to increase with increasing frequency. In the link budget calculations, we accounted for total losses of 8 dB which includes 3dB RF front-end loss and 5dB baseband and other losses. In practice, these losses may be higher and there may be other losses which will reduce achievable data rate, range or both. We didn't assume any NLOS losses as we assume there will always be a LOS available to the satellite. We will also need to take into account atmospheric absorption losses for uplink and downlink transmissions (inter-satellite links in space do not suffer from these losses) in particular for frequencies above 164GHz where water vapor absorption becomes significant. When these losses are present, we will need to either increase the transmit power or antenna gains to compensate for the losses.

**Table 2. Link budget analysis**

| Parameter | Value | Comments |
|---|---|---|
| Transmit Power | 33 dBm | Multiple PAs |
| Transmit Antenna Gain | 53 dBi | Element + array gain |
| Carrier Frequency | 100 GHz | Ref. for calculations |
| Distance | 1500 Km | LEO orbit |
| Propagation Loss | 195.92 dB | |
| Other path losses | 0 | Always LOS |
| Tx front end loss | 3.00 | Non-ideal RF |
| Receive Antenna Gain | 53.00 | Element + array gain |
| Received Power | -59.92 | |
| Bandwidth (BW) | 1 GHz | BW / comm-core |
| Thermal Noise PSD | -174 dBm/Hz | |
| Receiver Noise Figure | 5.00 dB | |
| Thermal Noise | -79 dBm | |
| SNR | 19.08 dB | |
| Implementation loss | 5 dB | Non-ideal Transceiver/ BB |
| Spectrum Efficiency (SE) | 4.73 b/s/Hz | |
| **Data rate / comm-core** | **4.73 Gb/s** | SE × BW |
| Number of comm-cores | 256 | BW and MIMO cores |
| **Aggregate data rate** | **1.21 Terabit/s** | 256×5.86 Gb/s |

## I. RF CIRCUITS & ANTENNAS

The design of electronic circuits at millimeter wave frequencies is challenging because the parasitic effects that are negligible at lower microwave frequencies become problematic at millimeter waves. Monolithic microwave integrated circuits (MMICs) designed using III-V semiconductor technologies, such as GaAs, GaN and InP, have been the technology of choice for millimeter wave applications. They offer superior performance compared to CMOS thanks to higher electron mobility, higher breakdown voltage, capabilities to handle high current densities, and the availability of high quality-factor passives. The III-V semiconductor-based commercial products from companies like Qorvo, Analog Devices and Avago are already offering power levels ranging from a few Watts to tens of Watts in the X (8-12 GHz), Ku (12-18 GHz), K (18-26 GHz) and Ka (26–40 GHz) bands of frequencies.

However, CMOS implementations promise higher levels of integration and reduced cost. Several developments in the past 10 years have combined to enable CMOS circuit blocks to operate at ever-increasing frequencies. Many commercial products based on CMOS implementation are now available in the market for the IEEE 802.11ad standard based access and point-to-point backhaul links in the 60GHz unlicensed-band. An extensive research is underway to extend CMOS technology to even higher frequencies. For example, a fully integrated E-band (both 71–76- and 81–86-GHz bands) power amplifier (PA) in 40-nm CMOS achieving a measured saturated output power of 20.9dBm with more than 15-GHz small-signal 3-dB bandwidth and 22% power-added efficiency (PAE) is recently disclosed in [8].

Another interesting technology that allows integration of digital logic and millimeter wave subsystems on a single chip is SiGe bipolar complementary metal oxide semiconductor (BiCMOS) technology. A 3-stage power amplifier (PA) with 3-dB bandwidth of 35 GHz (135-170 GHz) and implemented in 130 nm SiGe BiCMOS technology is presented in [9]. The PA offers saturated output power (Psat) from 5-8 dBm and the output referred 1 dB compression point (P1dB) varies from 1-6 dBm over the 135-170GHz frequency range. The nominal DC power consumption of this PA is 320 mW with peak PAE of 1.6%.

We notice that at higher millimeter waves, the device power levels and efficiencies are still low. However, we expect that improvements in power levels, efficiency and linearity for III-V semiconductor, SiGe BiCMOS and CMOS technologies will continue to happen in the coming years leading to viable solutions in the higher millimeter wave frequencies for the Space Internet application.

A key consideration for our proposal of Space Internet is to keep the cost low by using small form factor micro-satellites which are easier to launch and maneuver in orbit. Similarly, the ground and air-based platforms that communicate with satellites also needs to be low cost and easy to deploy and maintain. A key component of the system that determines the form factor is the size of the array which also has a significant impact on the transmitter EIRP as well as receiver sensitivity and thus directly impacts the system link budget. The antenna aperture size $A_e$ is related to the antenna gain or directivity ($D$) and the signal wavelength $\lambda$ as below [2]-[3]:

$$A_e = D\left(\frac{\lambda^2}{4\pi}\right)$$

It is obvious that to provide larger antenna gain, we need larger size antenna array. However, with increasing frequency (decreasing wavelength), antenna size goes down as square of wavelength as depicted in Figure 5. We note that with antenna array size of about $1 m^2$, we can achieve gain of up to 70dB at 150GHz frequency. For lower frequencies, we need to either increase the antenna array size beyond $1 m^2$ or expect a lower antenna gain. At around 30GHz, we achieve 50dB array gain with $1 m^2$ array size. We can also consider enhancing antenna arrays gain and beam patterns by using specially designed lenses combined with high performance feeds.

The smaller antenna size for the same gain makes higher millimeter wave frequencies attractive from form factor and weight perspective. We have noticed that at lower millimeter wave frequencies, higher power circuits providing up to several Watts output power are commercially available. However, at higher millimeter frequencies, technology is still evolving but currently available power levels are an order of magnitude lower. The lower power levels can partly be compensated by higher antenna gains at higher millimeter frequencies. However, as pointed out earlier, the atmospheric absorption losses becomes significant above 164GHz rendering these frequencies unattractive for satellite uplink and downlink transmissions where signals have to travel through the earth's atmosphere. Therefore, it may be prudent to initially limit the transmissions to/from the satellites to a maximum of 164GHz. The frequencies above 164GHz can be used for high-capacity inter-satellite links above earth's atmosphere.

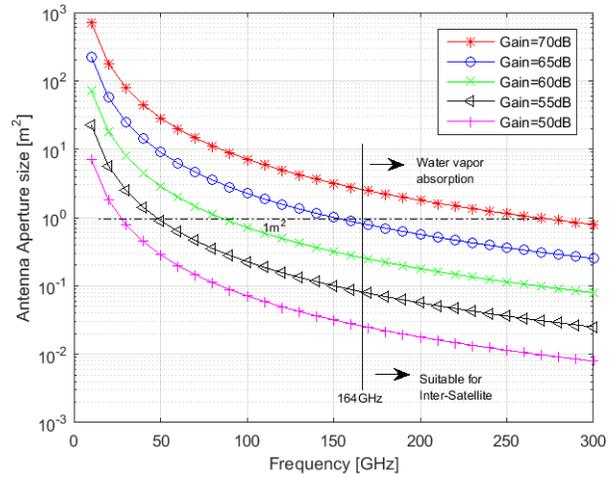

**Figure 5 Antenna aperture size as function of frequency**

## II. CONCLUSION

We outlined a vision of a Space Internet to make affordable Internet services available to everyone in the world via low-cost micro-satellites. We noted that about 4,600 such satellites operating at data rates in excess of Tb/s in LEO orbit can provide overall capacity of one Zetabyte/ month or 200GB/month for 5 Billion users Worldwide with signal latencies comparable to those offered by ground based systems. These ultra-high data rates are enabled by Multi-comm-core approach which enables using tens of GHz millimeter wave spectrum in the 10GHz to 275GHz frequency range. We identified a potential spectrum availability of over 100GHz which includes 57.75GHz for uplink, 56.2GHz for downlink and 38.75GHz for inter-satellite communication. With 5G mobile communications and next generation Wi-Fi expected to use millimeter wave spectrum, we can also develop a single wireless technology for satellite, cellular and Wi-Fi access as well as for back-haul communications to take advantage of the scale and further reduce costs. For example, the techniques developed for 5G millimeter wave mobile communication such as antenna-arrays based multi-beam communication, beam-steering and beam tracking can be applied to satellites in the LEO orbit which appear mobile to the ground based stationary and mobile systems and vice versa. We believe that our Space Internet proposal will bring us one step closer to connect and empower the whole humankind.

ACKNOWLEDGMENT

The author would like to thank his colleagues at SAMSUNG for valuable discussions and feedback.REFERENCES

[1] Cisco Visual Networking Index (VNI): Global Mobile Data Traffic Forecast Update, 2014-2019.
[2] F. Khan, Z. Pi, "Millimeter Wave mobile broadband (MMB): unleashing 3 – 300 GHz spectrum," IEEE WCNC, March 2011, Available: http://wcnc2011.ieee-wcnc.org/tut/t1.pdf.
[3] F. Khan, Z. Pi, "mmWave mobile broadband (MMB): unleashing the 3 – 300 GHz spectrum," in proc. of 34th IEEE Sarnoff Symposium, 3-4 May 2011, Princeton, New Jersey.
[4] Z. Pi, F. Khan, "An introduction to millimeter-wave mobile broadband systems," IEEE Communications Magazine, July 2011.
[5] W. Roh, et. al., "Millimeter-Wave Beamforming as an Enabling Technology for 5G Cellular Communications: Theoretical Feasibility and Prototype Results", IEEE Communicaitons Magazine, February 2014.
[6] F. Khan, "Architecting Terabit/s Wireless," submitted for publication.
[7] FCC, "Online Table of Frequency Allocation", July 2014.
[8] Dixian Zhao and Patrick Reynaert, "An E-Band Power Amplifier With Broadband Parallel-Series Power Combiner in 40-nm CMOS, IEEE Transactions on Microwave Thory and Techniques. February 2015.
[9] Neelanjan Sarmah, Bernd Heinemann, and Ullrich R. Pfeiffer, "A 135–170 GHz power amplifier in an advanced SiGe HBT technolog", IEEE Radio Frequency Integrated Circuits Symposium (RFIC), 2013.